\documentclass[12pt]{iopart}

\usepackage{times}
\usepackage{graphicx}

\input epsf.tex
\def\DESepsf(#1 width #2){\epsfxsize=#2 \epsfbox{#1}}
\def\NPB{{ Nucl. Phys.} B}

\def\PLB{{ Phys. Lett.}  B}
\def\PRL{ Phys. Rev. Lett.}
\def\PRD{{ Phys. Rev.} D}

\def\AP{ Astropart. Phys.}

\begin{document}
\title{Dark matter and Consequences of SUSY}
\author{ R. Arnowitt and B. Dutta\footnote{Present Address: Department of Physics, University of Regina, Regina
SK, S4S  0A2 Canada}}
\address{Center For Theoretical Physics,
Department of Physics, Texas A\&M University, College
Station, TX 77843-4242, USA}
\begin{abstract}
We examine here the constraints from the amount of relic density of
neutralino dark matter and other experiments have on the SUSY parameter
space for the mSUGRA model and for models with non-universal soft breaking
at the GUT scale. In mSUGRA, the allowed amount of dark matter restricts
the SUSY parameter space to a narrow band in $m_0 - m_{1/2}$ (except at very
large $\tan\beta$). The Higgs mass and $b\rightarrow s\gamma$ constraints produce a lower
bound of $m_{1/2} \stackrel{>}{\sim}$ 300GeV and if the muon magnetic moment anomaly can be
interpreted as a 3$\sigma$ deviation from the  Standard Model, one also
obtains an upper bound of $m_{1/2} \stackrel{<}{\sim}900$GeV, making the SUSY spectrum well
accessible to the LHC. The $B_s\rightarrow \mu \mu$ decay is seen to be accessible to
the Tevatron Run2B with $15$ fb$^{-1}$ for $\tan\beta\stackrel{>}{\sim}30$. However, only parts of
the spectrum will be accessible to the NLC if it's energy is below 800GeV.
Non-universal soft breaking opens new regions of parameter space. Thus the
$m_{1/2}$ lower bound constraint of $b\rightarrow s\gamma$ and also the Higgs mass can be
reduced greatly if the gluino mass is assumed larger at the GUT scale
(allowing for a lighter gaugino spectrum), and non-universal Higgs soft
breaking masses at the GUT scale can open new allowed regions at relatively
low $m_{1/2}$ and high $m_0$ where dark matter detection cross sections may be
increased by a factor of ten or more.

\end{abstract}

\section{Introduction}

Supersymmetry is a natural solution to the gauge hierarchy problem that
sets in for the Standard Model (SM) at the TeV scale. Thus if one
supersymmetrizes the Standard Model particle spectra, one can build a model
going past the TeV scale yet consistent with all the successes of the
Standard Model below 1 TeV. If one continues such models to yet higher
energies, one find the remarkable grand unification of the three gauge
coupling constants at the GUT scale $M_G\cong 2\times10^{16}$ GeV, a result
consistent with the LEP data at the percent level. Theoretical models which
yield this unification arise naturally in supergravity (SUGRA) grand
unification\cite{sugra1,sugra2}, and such models with R-parity invariance have the additional
remarkable feature of predicting the existence of cold dark matter
(CDM)\cite{gold,ellis0},
the lightest neutralino $\tilde\chi^0_1$, with a relic density amount comparable to
what is observed.

We discuss here some of the other consequences that might be expected of
such SUGRA models. Already existing experiments have begun to restrict the
SUSY parameter space significantly. Most significant of these are the
amount of CDM, the Higgs mass bound, the $b\rightarrow s \gamma$  branching ratio,
and (possibly) the muon $a_\mu$ anomaly. We start the discussion with the
simplest model, mSUGRA, with universal soft breaking parameters, and
discuss additional signatures of SUSY that might be seen at accelerators,
i.e. the $B_s\rightarrow \mu^+ +\mu^-$ decay at the Tevatron, and the processes 
$e^+ +e^- \rightarrow \tilde\tau_1^++ \tilde\tau_1^-$ or 
$e^+ + e^- \rightarrow \tilde\chi^0_1 + \tilde\chi^0_2$ at future linear colliders
(LC).  We will then consider non-universal models (with non-universal
gaugino masses and non universal Higgs soft breaking masses at $M_G$) to see
how robust the mSUGRA predictions are, and where non-universal soft
breaking signals might reside.

\section{mSUGRA Models}

The mSUGRA models\cite{sugra1,sugra2} depend on four extra parameters and one sign (and as
such is the most predictive of the SUSY models). We take these to be $m_0$
(the universal scalar soft breaking mass at $M_G$), $m_{1/2}$ (the universal
gaugino masses as $M_G$), $A_0$ (the cubic soft breaking mass at $M_G$), 
$\tan\beta = <H_2>/<H_1>$ (at the electroweak scale), and the sign of the Higgs mixing
parameter $\mu$ (which appears in the superpotential W as  $\mu
H_1 H_2$). We
examine the parameter range $m_0 >0$, $m_{1/2}\leq$1 TeV (which corresponds to the
LHC reach of $m_{\tilde g} < 2.5$ TeV), $2 < \tan\beta < 55$ and $|A_0| <  4 m_{1/2}$.

mSUGRA makes predictions about two items involving dark matter: the
neutralino-nucleus cross section being looked for by terrestrial detectors of dark
matter from the Milky Way halo, and the mean amount of relic dark matter in
the universe left over from the Big Bang. For detectors with heavy nuclei
in their targets, the spin independent $\tilde\chi^0_1$-nucleus cross section dominates,
which allows one to extract the $\tilde\chi^0_1-$proton cross section 
$\sigma_{\tilde\chi^0_1-p}$. The
basic quark diagrams here are the scattering through s-channel squarks, and
t-channel CP even Higgs bosons, $h$ and $H$. For the relic density analysis,
one must calculate the neutralino annihilation cross section in the early
universe which proceeds through s channel $Z$ and Higgs poles ($h$, $H$, $A$) and
t-channel sfermions. However, if a second particle becomes nearly
degenerate with the $\tilde\chi^0_1$, one must include it in the annihilation channels
which leads to the phenomena of co-annihilation. In SUGRA models, this
accidental near degeneracy occurs naturally for the light stau, $\tilde\tau_1$. One
can see this qualitatively for low and intermediate tanbeta where one can
solve the RGE analytically. Thus for the right selectron, $e_R$, one finds
\begin{eqnarray}
                  m_{eR}^2 &=&m_0^2 +0.15m_{1/2}^2 - sin^2\theta_W M_W^2
cos2\beta,\\
                   M_{\tilde\chi^0_1}^2 &=&0.16 m_{1/2}^2
\end{eqnarray}
For $m_0 = 0$,  the two become degenerate at $m_{1/2}\cong350$ GeV, and
co-annihilation thus begins at $m_{1/2}\cong350$ GeV (more precisely for
the $\tilde\tau_1$ which is the lightest slepton). As $m_{1/2}$ increases, 
$m_0$ must
increase (to keep the $\tilde\tau_1$ heavier than the $\tilde\chi^0_1$) and one gets narrow
bands of allowed regions in the $m_0 - m_{1/2}$ plane for each $\tan\beta$ and 
$A_0$.

One starts the analysis at the $M_G$ and uses the renormalization group equations
(RGE) to obtain predictions at the electroweak scale. In carrying out these
calculation, it is necessary to include a number of corrections and we
list some of these here:
(1) We use two loop gauge and one loop Yukawa RGE in running from $M_G$ to the
electroweak scale $M_{\rm EW}$, and three loop QCD RGE below $M_{\rm EW}$  for light quark
contributions.
(2) Two loop and pole mass corrections are included in the calculation of $m_h$.
(3) One loop correction to $m_b$ and $m_\tau$ are included\cite{rattazi}
(4) All stau-neutralino co-annihilation channels are included in the relic
density calculation \cite{bdutta,ellis,gomez}. (Chargino-neutralino co-annihilation does
not occur for $m_{1/2} < 1$ TeV.) Large $\tan\beta$ NLO SUSY corrections to
$b\rightarrow s\gamma$ are included\cite{degrassi,carena2,isidori}.
We do not include Yukawa unifications or proton decay constraints as these
depend sensitively on post-GUT physics, about which little is known.

\begin{figure}\vspace{-0cm}
\centerline{ \DESepsf(figadkm.epsf width 8 cm) }
\caption {\label{fig1}  Example of a leading contribution to the decay $B_s\rightarrow \mu \mu$. Each
vertex with a dot has a factor of tanbeta so that the diagram is
proportional to $\tan\beta^3$.} 
\end{figure}

 \begin{figure}\vspace{-0cm}
 \centerline{ \DESepsf(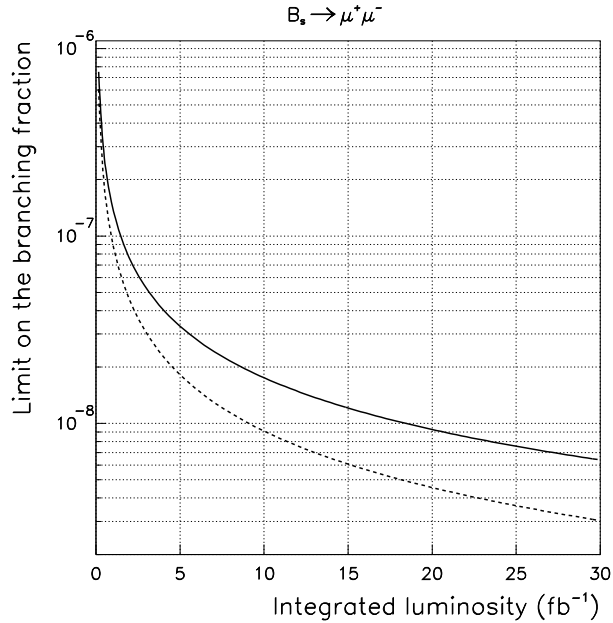 width 8 cm) } 
\caption {\label{fig2}  Expected sensitivity to $B_s\rightarrow \mu \mu$ of the CDF detector as a
function of luminosity. The solid curve is a conservative estimate, and the
dotted curve gives a limiting sensitivity.[20]}
\end{figure} 

\begin{figure}
 \centerline{ \DESepsf(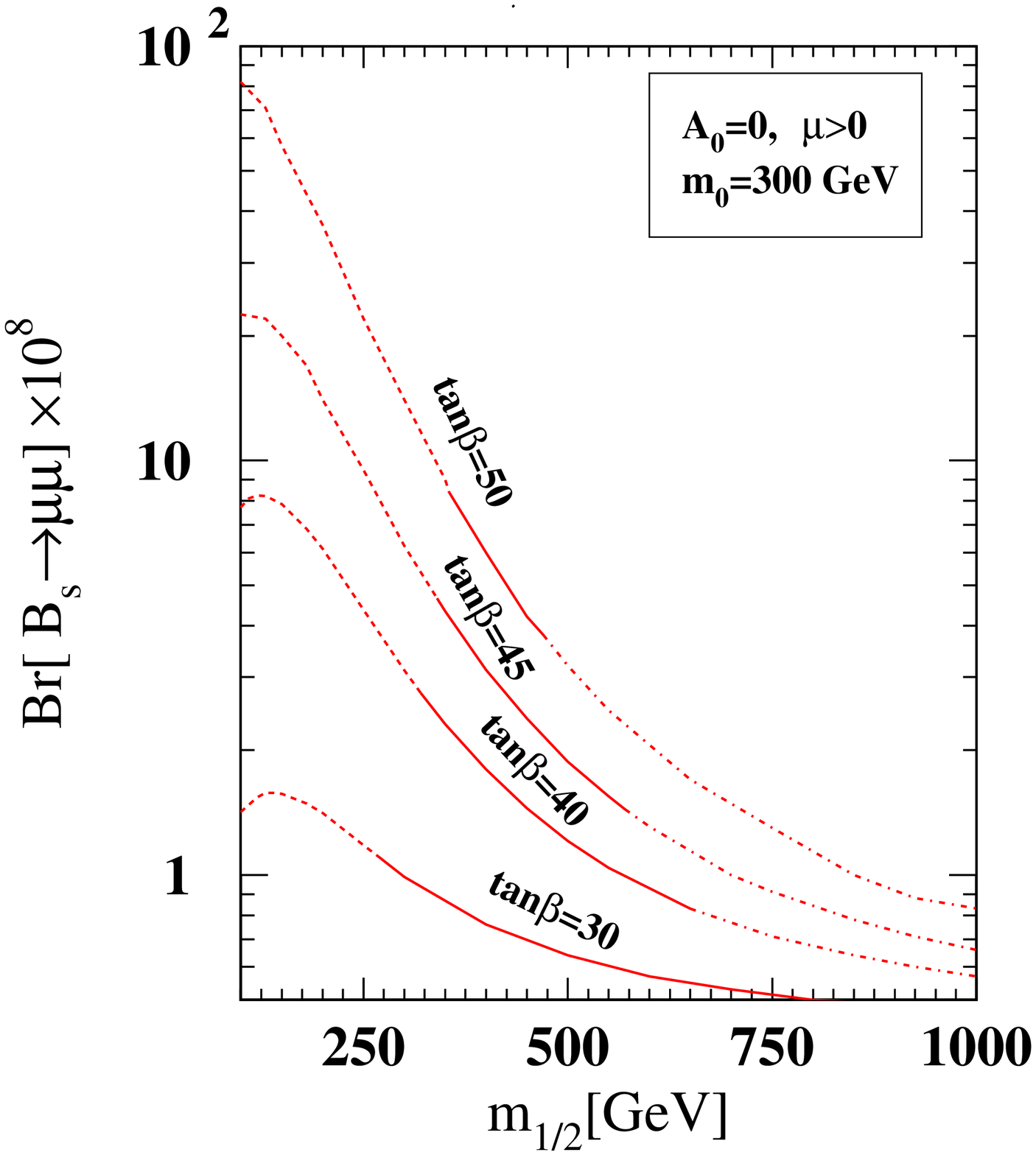 width 8 cm) } 
\caption {\label{fig3}  mSUGRA predicted values of B($B_s\rightarrow \mu \mu$) for different values of
tanbeta for $A_0 =0$ and $m_0 = 300$.}
\end{figure}

We use the following experimental input to constrain the SUSY parameter space:
(1) Global fits for the CMB and other astronomical measurements now
restrict the amount of CDM considerably\cite{turner}. We use here the range:
\begin{equation}
 0.07 < \Omega_{\rm CDM} h^2 <0.21
\label{om}\end{equation}

The MAP data, due out perhaps early next year should be able to restrict
this range considerably.

(2) The LEP lower bound $m_h > 114$ GeV\cite{higgs1}, is a significant constraint for
$\tan\beta\stackrel{<}{\sim}30$, and in fact an increase of three to five GeV would cover
most of the parameter space. Unfortunately, however, the theoretical
evaluation of $ m_h$ still has an error of $\sim(2-3)$GeV, and so we interpret the
LEP bound to mean $(m_h)^{theory}> 111$ GeV.

(3) The CLEO data for the decay $b\rightarrow s \gamma$\cite{bsgamma} has both systematic and
theoretical error, and so we use a relatively broad range for the branching
ratio around the CLEO central data:
\begin{equation} 1.8\times10^{-4} < B(B \rightarrow X_s \gamma) <
4.5\times10^{-4}
\label{bs}
\end{equation} 
The  $b\rightarrow s \gamma$ rate produces a significant constraint for large tanbeta.

(4) Shortly after this conference, the Brookhaven E821 experiment published
new results of their measurement of the muon anomaly\cite{BNL} which reduced the
statistical error by a factor of 2 (and the systematic error somewhat).
Also new more accurate data from CMD-2, BES, ALEPH and CLEO have allowed a
more accurate determination of the SM contribution to $a_\mu$, and there have
been two new evaluations of this \cite{dav,hag} (See also talk by T. Teubner at
this conference). If the $e^+ - e^-$ data is used to calculate the SM
contribution, both groups find a 3$\sigma$ deviation between experiment and
the SM prediction e.g.\cite{dav}
\begin{equation}  
\Delta a_\mu = 33.9 (11.2)\times10^{-10}\label{g2}
\end{equation}
On the other hand if the tau data of ALEPH and CLEO are used (with
appropriate CVC breakdown corrections) the deviation from the SM is reduced
to only 1.6$\sigma$ and the two evaluations are statistically inconsistent
\cite{dav}. The matter remains unclear as to which result is correct. In order to
see the significance of Eq. (5), we will here use this result, and assume a
2$\sigma$ lower bound on delta $a_\mu$ exists and that this is attributable to
SUSY. SUSY, in fact makes a significant contribution to $a_\mu$, and
if $a_{\mu}^{\rm SUSY}$ were $\stackrel{<}{\sim}10^{-10}$, the squark and gluino mass spectrum would be
pushed into the TeV domain.

The combination of the $m_h$(for low $\tan\beta$ ) and $b\rightarrow s \gamma$ (for high
$\tan\beta$) constraints produces a lower bound on $m_{1/2}$ over the entire
parameter space of $m_{1/2} \stackrel{>}{\sim}$(300-400)GeV, and consequently 
$m_{\tilde\chi^0_1}  \stackrel{>}{\sim}$(120 -
160)GeV. This means that most of the parmaeter space is in the 
$\tilde\tau_1-\tilde\chi^0_1$ co-annihilation domain, and hence in order to satisfy the CDM amount,
$m_0$ is approximately determined by $m_{1/2}$ (for fixed tanbeta, $A_0$). The sign of
$\mu$ and $\Delta a_\mu$ are correlated\cite{nano,chat}, and so the assumption of Eq. (5)
means that $\mu > 0$.
\begin{figure}\vspace{-0cm}
 \centerline{ \DESepsf(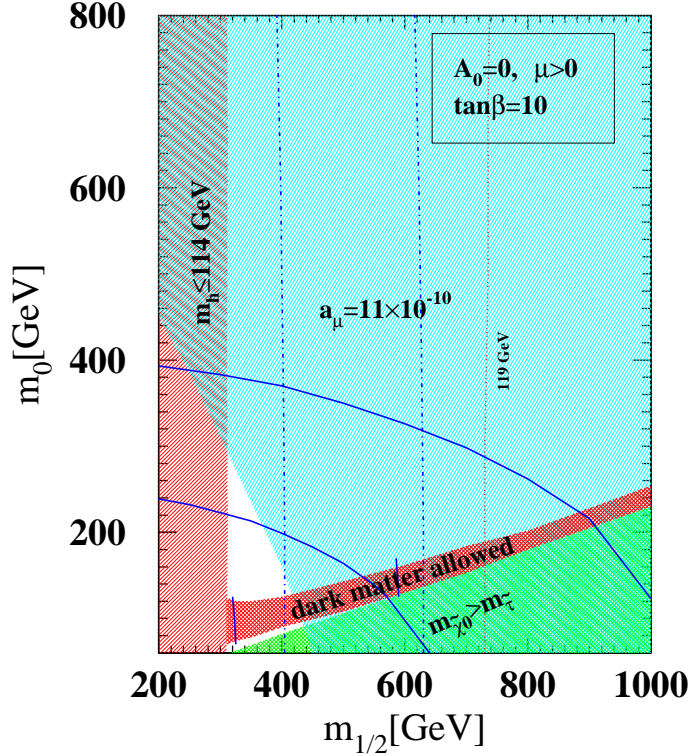 width 8 cm) } 
\caption {\label{fig4}  mSUGRA allowed region for $\tan\beta = 10$, $A_0 = 0$, $\mu >0$. The shaded
upper right region is forbidden by the $a_\mu$ bound at the $2\sigma$ level if
Eq. (5) is valid. The dot-dash lines are for the LC $\tilde\chi^0_1 - \tilde\chi^0_2$ signal for the
500GeV machine (left line) and 800GeV machine (right line). (The LC is
sensitive to regions to left of the line.) The curved solid lines are for
the LC $\tilde\tau_1-\tilde\tau_1$ signal (the lower one for the 500 GeV machine and the
higher one for the 800GeV machine). The short vertical lines are for DM
detection cross sections, $\sigma_{\tilde\chi^0_1-p} = 5\times 10^{-9}$pb (left line) and
$1\times10^{-9}$pb (right).}
\end{figure}

\begin{figure}\vspace{-0cm}
 \centerline{ \DESepsf(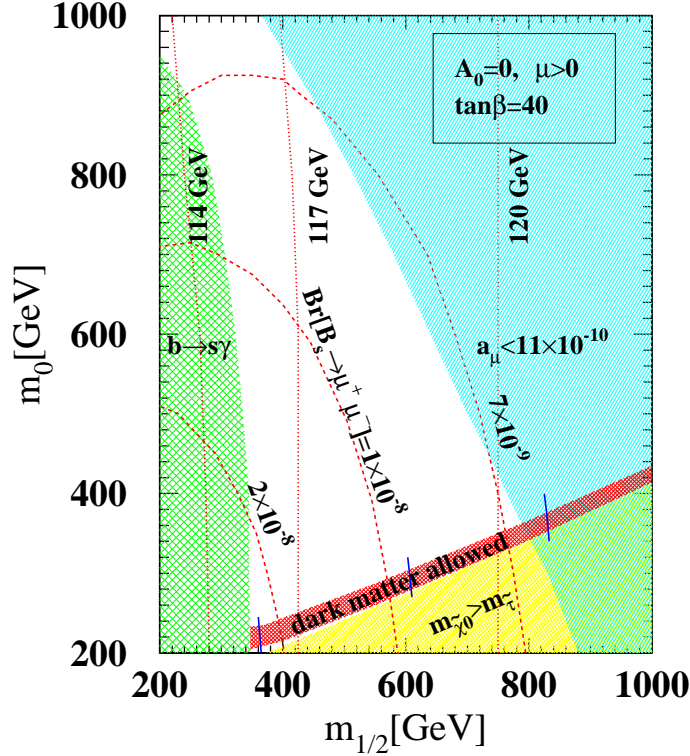 width 8 cm) } 
\caption {\label{fig5}  The same as Fig. 4 for $\tan\beta = 40$, $A_0 = 0$, the dashed lines giving
contours of $B_s\rightarrow \mu \mu$ branching ratios. The lower short vertical line is
$\sigma_{\tilde\chi^0_1-p} =3\times10^{-8}$pb and the upper one is $1\times10^{-9}$pb}
\end{figure}

\begin{figure}\vspace{-0cm}
 \centerline{ \DESepsf(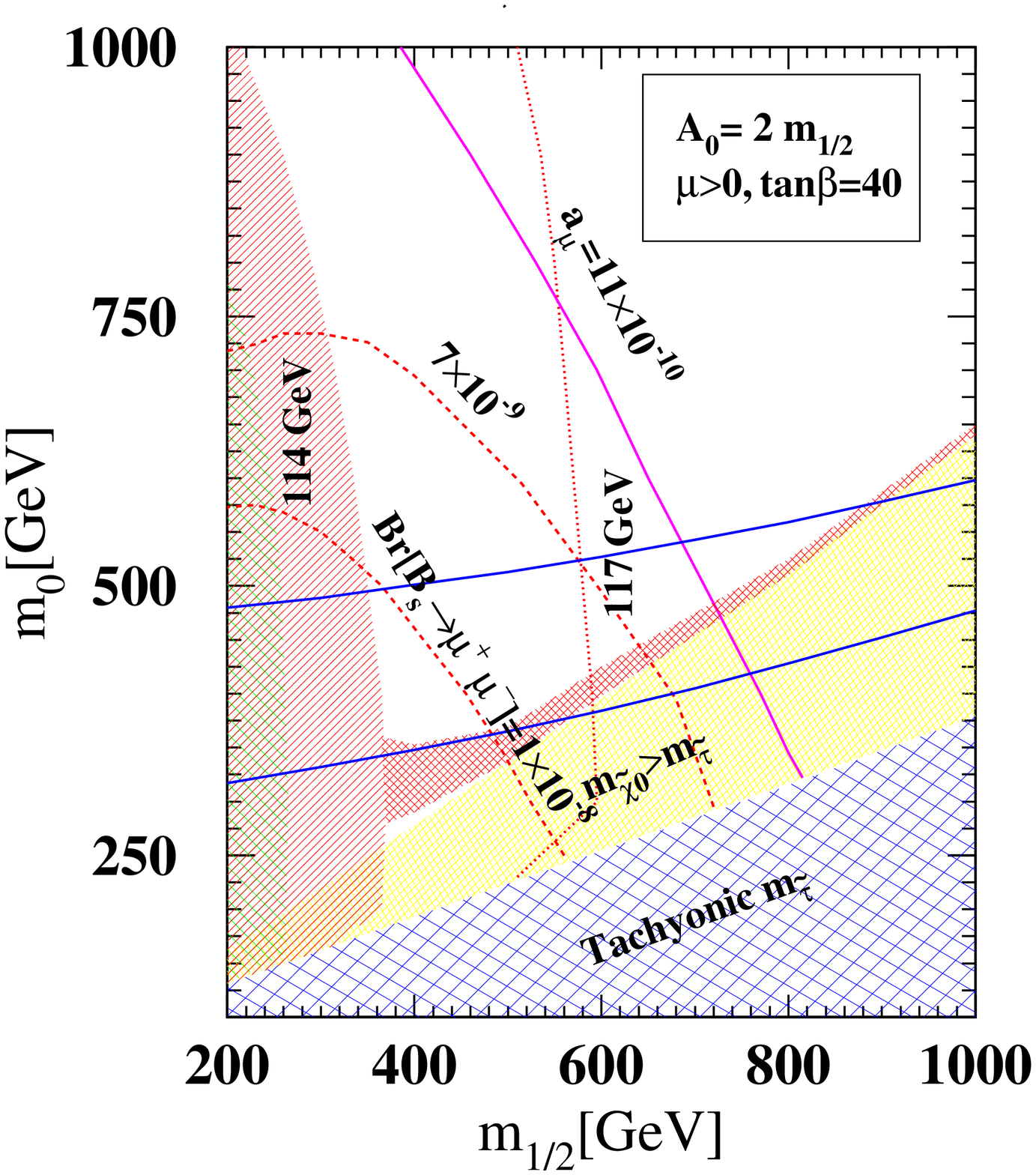 width 8 cm) } 
\caption {\label{fig6}  Same as Fig. 5 for $\tan\beta = 40$, $A_0 = 2m_{1/2}$, $\mu >0$.}
\end{figure}

\begin{figure}\vspace{-0cm}
 \centerline{ \DESepsf(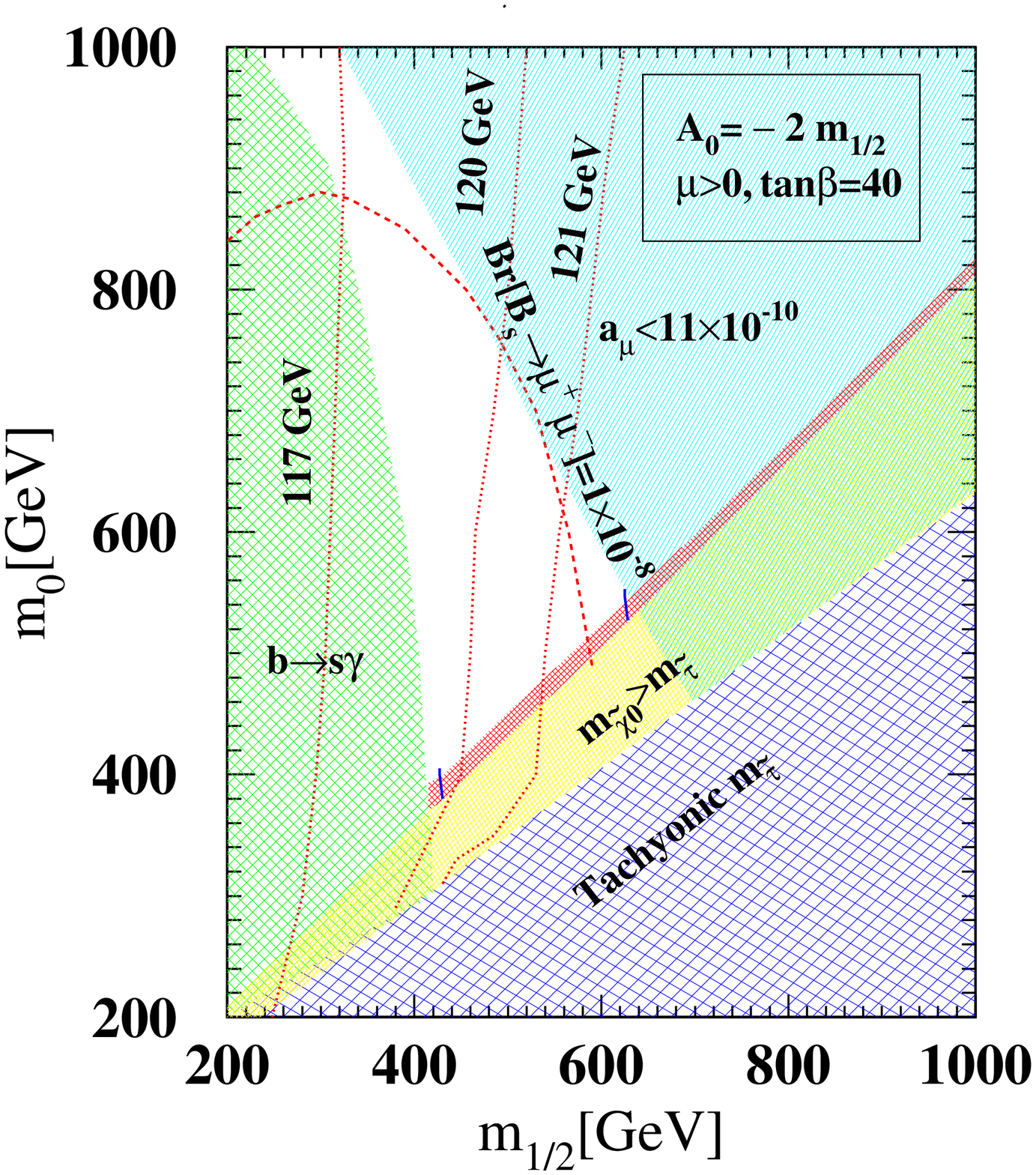 width 8 cm) } 
\caption {\label{fig7}  Same as Fig. 5 for $\tan\beta = 40$, $A_0 = - 2 m_{1/2}$, $\mu >0$}
\end{figure}

\begin{figure}\vspace{-0cm}
 \centerline{ \DESepsf(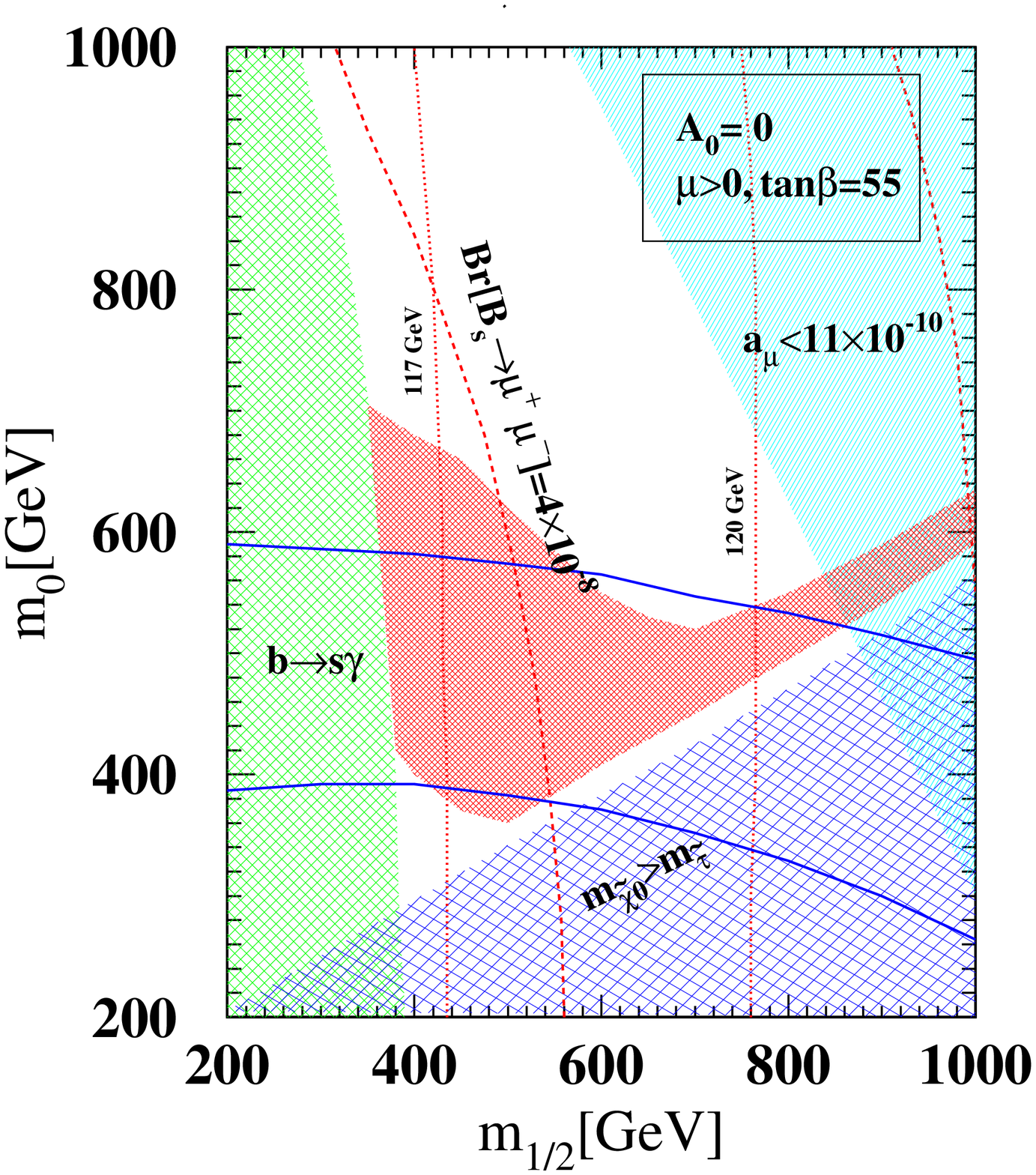 width 8 cm) } 
\caption {\label{fig8}Same as Fig. 5 for $\tan\beta = 55$, $A_0 =0$, $\mu >0$ and $m_t = 175$GeV,
$m_b= 4.25$. (Note that the results are sensitive to the exact values of $m_t$ and
$m_b$ used}
\end{figure}
\section{SUSY Signals At The Tevatron and Linear Colliders}

Before examining in more detail the effects of existing experiments on the
SUSY parameter space, we discuss the $B_s\rightarrow \mu \mu$ decay at the Tevatron and
also examine SUSY signals that might survive at a linear collider.

The $B_s\rightarrow \mu \mu$ opens an additional window at the Tevatron for
investigating the mSUGRA parameter space\cite{bobeth,bdutta3,dedes,tata}. The SM predicts a
very small branching ratio of B[$B_s\rightarrow \mu \mu$] = $(3.1 \pm
1.4)\times10^{-9}$.
Further, the SUSY contribution can become quite large for large tanbeta, as
the leading diagrams (an example is given in Fig. 1) grow like $\tan\beta^3$
and hence the branching ratio grows as $\tan\beta^6$. A set of cuts have been
obtained eliminating background (mostly gluon splitting $g \rightarrow bb$ and fakes)
\cite{bdutta3} leading to a sensitivity of B[$B_s\rightarrow \mu \mu$] $\stackrel{>}{\sim}
6\times10^{-9}$ for  15fb$^{-1}$ 
/detector. Fig. 2 shows the sensitivity (for a single detector) as a
function of luminosity, and Fig. 3 shows the expected mSUGRA branching
ratio for different $\tan\beta$ for the example of $A_0 = 0$, $m_0 = 300$GeV. One see
that the Tevatron will be sensitive to this decay for $\tan\beta \stackrel{>}{\sim}30$.

It is interesting to examine what possible SUSY signals of mSUGRA a linear
collider might see. We consider here two possibilities: $\sqrt{s} = 500$GeV
and $\sqrt{s} = 800$GeV. We've seen above that the $m_h$ and 
$b\rightarrow s\gamma$
constraint already mean that $m_{1/2} > (350 - 400)$ GeV,
and so for mSUGRA this means that gluinos and squarks would generally be
beyond the reach of such machines (as well as selectrons and smuons for a
large part of the parameter space). The most favorable SUSY signals are then
\begin{eqnarray}
            e^+ + e^- &\rightarrow& \tilde\chi^0_2 + \tilde\chi^0_1 \rightarrow  
	    (l^+ + l^- +\tilde\chi^0_1) +\tilde\chi^0_1 \\\nonumber 
	    e^+ + e^- &\rightarrow& \tilde\tau_1^+ + \tilde\tau_1^- \rightarrow 
( \tau +\tilde\chi^0_1) + ( \tau +\tilde\chi^0_1)
\end{eqnarray} 
where $l^+$ or $l^-$ means any charged lepton. Since for mSUGRA one has that
$m_{\tilde\chi^0_2} \simeq 2m_{\tilde\chi^0_1}$, the mass reach for these particle are
\begin{equation}1/2m_{\tilde\chi^0_2} \simeq m_{\tilde\chi^0_1}
\stackrel{<}{\sim}165 (265)\end{equation}
\begin{equation}m_{\tilde\tau_1}\stackrel{<}{\sim}250 (400)$ GeV for $\sqrt s = 500 (800)
\end{equation}

There are a number of SM backgrounds that have to be considered to see if a
clean signal remains. For example, WW production with decay into tau final
states can be controlled by polarizing the beams. Cuts to eliminate other
possible backgrounds are under consideration\cite{bdutta4}.

\section{The mSUGRA Parameter Space}

We now summarize the effects of the constraints from $m_h$, $b\rightarrow s\gamma$, dark
matter density and $a_\mu$ on the mSUGRA parameter space, and also what might
be expected from the $B_s\rightarrow \mu \mu$ observation at the Tevatron and linear
collider (LC) signals. Fig. 4 exhibits the parameter space for $\tan\beta =
10$, $A_0 = 0$. One sees that the $a_\mu$ bound (if valid) combined with the 
$m_h$
bound and the relic density constraint leaves very little parameter space
at low $\tan\beta$.  Either of the two linear collider signals could scan the
full remaining space. The dark matter detection cross sections are of size
that would be observable by the planned future DM detectors. Figs. 5,6,7
for $\tan\beta = 40$, $A_0 = 0$, $2m_{1/2}$ and -$2m_{1/2}$ exhibit a larger allowed
parameter space. The Tevatron's $B_s\rightarrow \mu \mu$ covers the full allowed region
for $A_0 = 0$ and - $2 m_{1/2}$, and about half the space for $A_0 = 2 m_{1/2}$. In each
case, the NLC at 500GeV can only cover a part of the parameter space. We
note that the stau-stau LC signal extends to high $m_{1/2}$ and relatively low
$m_0$, while the $\tilde\chi^0_1 - \tilde\chi^0_2$ signal can cover large $m_0$ but limited $m_{1/2}$. In view
of the nature of the allowed DM channel, the former is of more
significance. At very high $\tan\beta$, a bulge in the DM allowed channel
develops at low $m_{1/2}$ due to the fact the heavy Higgs ($A$,$H$) become light,
allowing a rapid early universe annihilation through the $A$ and $H$ s-channel
diagrams. This is shown most dramatically in Fig. 8 for $\tan\beta = 55$, $A_0 =
0$. Here even the 800 GeV LC cannot cover the full parameter space, though
the Tevatron signal of $B_s\rightarrow \mu \mu$ would be observable over the full
parameter space.

\section{Non-Universal Models}

We examine next some SUGRA models with non-universal soft breaking  to see
what aspects of the results of Sec. 4 are maintained. We consider
specifically the case of non-universal gaugino masses and non-universal
Higgs soft breaking masses at MG. Two striking effects in mSUGRA are the
$\tilde\chi^0_1-\tilde\tau_1$ co-annihilation channel which leads to a narrow band  in the $m_{1/2}-
m_0$ plane of allowed relic density, and the fact that the combined effects
of $b\rightarrow s\gamma$ and the $m_h$ bounds require 
$m_{\tilde\chi^0_1} \stackrel{>}{\sim} 120$ (i.e. $m_{1/2}\stackrel{>}{\sim} 300$GeV).

The possibility of non-universal gaugino masses at $M_G$ can relax
significantly the constraints of $b\rightarrow s\gamma$ and $m_h$. Thus if the gluino mass
is increased, it effects the stop mass which has major effects on both
$b\rightarrow s\gamma$ and $m_h$. For example, a gluino mass which is twice the universal
value at $M_G$ reduces the lower bound on $m_{\tilde\chi^0_1}$ to 75GeV 
($m_{1/2}\stackrel{>}{\sim} 190$GeV) and
also $m_{\tilde\chi^\pm_1}$ to $\sim150$GeV, making these particles more accessible. On the
other hand, the near degeneracy between $\tilde\tau_1$ and $\tilde\chi^0_1$ remains, since this
depends mainly on the U(1) gaugino mass $\tilde m_1$.
Non-universality in the Higgs masses at $M_G$, i. e. $m_{H_{1,2}}^2 = m_0^2 (1 +
\delta_{1,2})$, can also produce new effects. While this introduces  two
additional parameters into the model, one can understand their effects
since $\mu^2$ controls much of the physics. Thus if $\mu^2$ decreases, the
Higgsino content of the neutralino increases and this has two effects: it
increases the $\tilde\chi^0_1 - \tilde\chi^0_1 -Z$ coupling and also 
$\sigma_{\tilde\chi^0_1-p}$. To see what effects occur
qualitatively, we note that for small and intermediate $\tan\beta$ the RGE can
be solved analytically and give
\begin{equation} 
            \mu^2 = (\mu^2)_{\rm mSUGRA} + t^2/(t^2 - 1) [ - 1/2(1+D_0) \delta_2 +
\delta_1/t^{2}]m_0^2 
\end{equation}
where $t = \tan\beta$, $D_0\cong1 - (m_t/200sin\beta)^2\cong0.25$. A positive
$\delta_2$
then decreases $\mu^2$ while $\mu^2$ is relatively insensitive to $\delta_1$. Fig. 9
exhibits the allowed relic density regions for the case of $\delta_2 = 1$,
$\tan\beta = 40$, $A_0 =m_{1/2}$. One sees that a new region of allowed relic density
arises for small $m_{1/2}$ and large $m_0$ from the increased annihilation through
the $Z$ s-channel. (The usual $\tilde\tau_1 - \tilde\chi^0_1$ co-annihilation region is present
also.) This region is mostly unobservable at a 500GeV NLC. Fig. 10 shows
the corresponding DM cross sections. The Z-channel region can increase
$\sigma_{\tilde\chi^0_1-p}$ by a factor of 10 or more.

\begin{figure}\vspace{-0cm}
 \centerline{ \DESepsf( 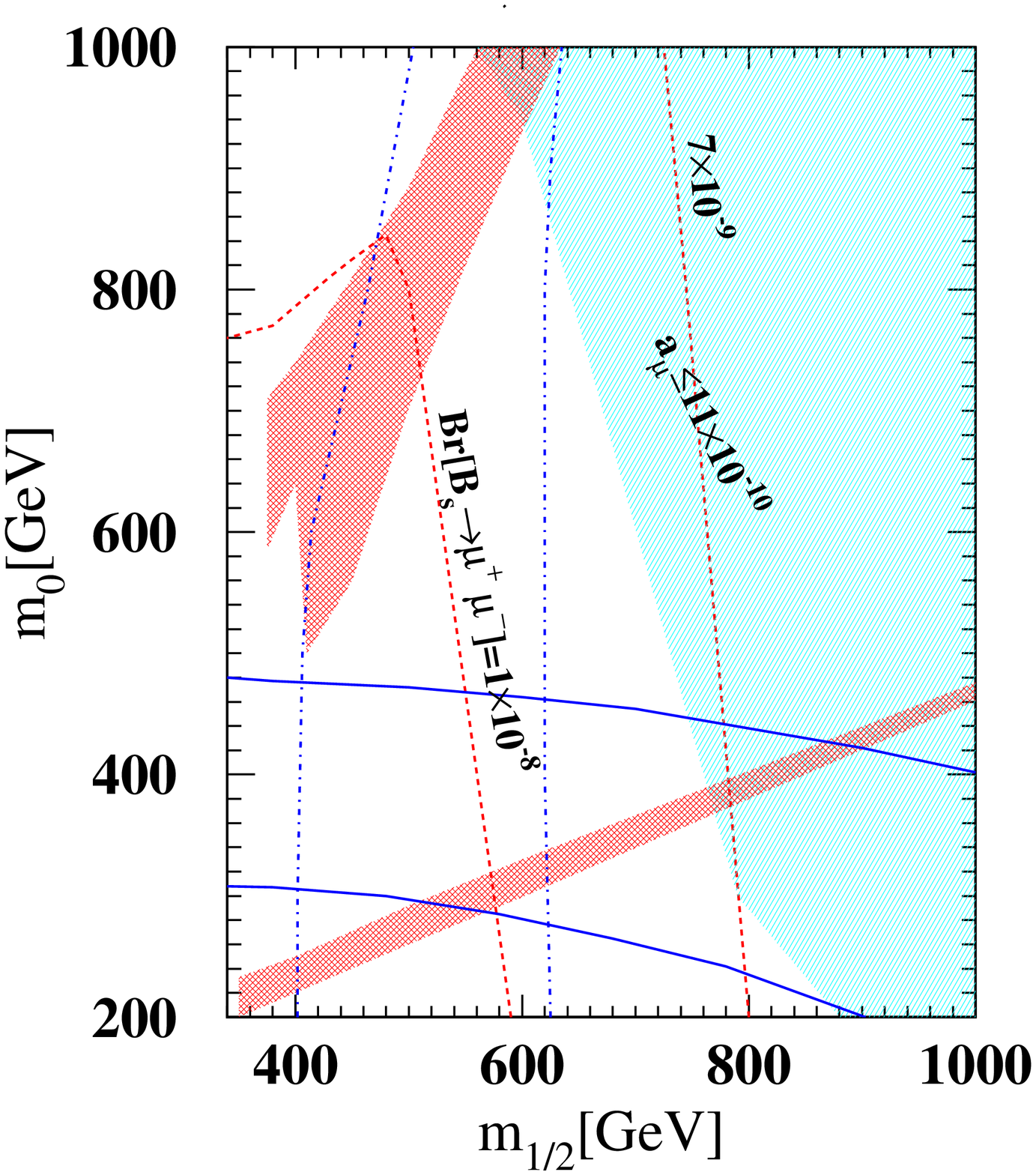 width 8 cm) } 
\caption {\label{fig9} Same as Fig. 5 for $\delta_2 = 1$, $\tan\beta = 40$, $A_0
= m_{1/2}$, $\mu >0$. The
shaded region for large $m_0$ and low $m_{1/2}$ is due to the increased
annihilation through the Z-channel. The lower stau-neutralino narrow band
is essentially unchanged.}
\end{figure} 

\begin{figure}
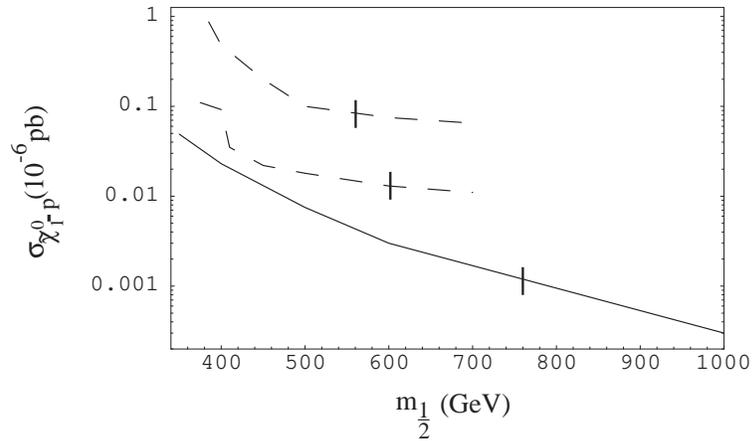
\vspace{-0cm}
 \centerline{ \DESepsf(adhs4.epsf width 10 cm) } 
\caption {\label{fig10}Dark matter detection cross section for $\tan\beta = 40$, $A_0 = m_{1/2}$,
$\mu >0$, $\delta_2 = 1$. The upper dashed band is the new region due to the
increased Z-channel annihilation. The lower line us the upper value from
the narrow $\tilde\tau_1 - \tilde\chi^0_1$ co-annihilation channel. The short
vertical lines are the 2$\sigma$ bound from the $a_{\mu}$ anomaly.}
\end{figure}

\section{Conclusions and Summary}

We have discussed here the current restrictions on the SUSY parameter
space from existing experiments and what may be obtained from future
experiments at the Tevatron (from the  $B_s\rightarrow \mu \mu$ decay) and from linear
colliders (using $\tilde\tau_1 -\tilde\tau_1$ and $\tilde\chi^0_1 - \tilde\chi^0_2$ signals). We have examined
mSUGRA models and also non-universal SUGRA models.

For mSUGRA one finds the $\tilde\tau_1 - \tilde\chi^0_1$ co-annihilation limits the allowed
parameter space to a narrow band except for large $\tan\beta$. The combined $b\rightarrow s\gamma$ and $m_h$ bounds then requires $m_{\tilde\chi^0_1} \stackrel{>}{\sim} $120GeV, forbidding a light
neutralino. The Tevatron with 15 fb$^{-1}$/detector luminosity could scan
almost the entire parameter space for $\tan\beta \stackrel{>}{\sim}40$ (if
$a_{\mu}^{\rm SUSY} \stackrel{>}{\sim} 10\times 10^{10}$) 
using the $B_s\rightarrow \mu \mu$ decay. A linear collider at 800GeV can
scan almost all the parameter space, though a 500 GeV machine would miss
much of the parameter space for large $\tan\beta$. Dark matter detection cross
sections are all within the range of sensitivity for planned detectors
(i.e. $\stackrel{>}{\sim} 10^{-10}$ pb).

Non-universal models introduce new phenomena. Thus a non-universal gluino
mass can greatly weaken the lower bound on $m_{\tilde\chi^0_1}$ due to
$b\rightarrow s\gamma$ and $m_h$.
An increased non-universal $H_2$ Higgs mass at $M_G$ can give rise to a new
region of allowed relic density for low $m_{1/2}$ and large $m_0$ from rapid $Z$-
channel annihilation, while maintaining the usual $\tilde\tau_1 - \tilde\chi^0_1$ co-annihilation
channel. The dark matter cross sections can increase by a factor of 10 or
more in the new $Z$-channel region.

\section{Acknowledegement}
 This work was supported in part by the National Science
Foundation Grant PHY - 0101015.

\end{document}